# Growth and Properties of Quaternary Alloy Magnetic Semiconductor (InGaMn)As

Shinobu Ohya[1], Hiromasa Shimizu[1], Yutaka Higo[1], Jiaming Sun[1], and Masaaki Tanaka[1,2]

*1) Department of Electronic Engineering, The University of Tokyo, 7-3-1 Hongo, Bunkyo-ku, Tokyo, 113-8656, Japan.*
*2) CREST, Japan Science and Technology Corporation, 4-1-8 Honcho, Kawaguchi, 332-0012, Japan.*

We have studied growth and properties of quaternary alloy magnetic semiconductor (InGaMn)As grown both on GaAs substrates and on InP substrates by low-temperature molecular-beam epitaxy (LT-MBE). (InGaMn)As thin films were ferromagnetic below ~30 K, exhibiting strong magneto-optical effect. The lattice constant of $[(In_yGa_{1-y})_{1-x}Mn_x]As$, whose Mn concentration $x$ is below 4%, is slightly smaller than that of $In_yGa_{1-y}As$ with the same In/Ga content ratio. We have also obtained very smooth surface morphology of nearly lattice matched (InGaMn)As thin films grown on InP substrates, which is important for application to thin-film type magneto-optical devices integrated with III-V opto-electronics.

A variety of II-VI and III-V based magnetic semiconductors, such as (CdMn)Te,[1] (InMn)As[2] and (GaMn)As,[3,4] have been extensively studied in the past years, because they have both properties of semiconductors and of magnetic materials and their spin-related functions could be used for novel "spintronic" devices. Among them, feasibility of growing magnetic quantum heterostructures based on (GaMn)As can give a new degree of freedom in the design of semiconductor materials and devices.[3-7] One of the recent achievements are, for example, very high tunneling magneto-resistance (TMR) ratios over 70% (maximum 75%) observed in (GaMn)As/AlAs/(GaMn)As heterostructures.[7] All of these studies mentioned above were done on *ternary* alloy magnetic semiconductors.

In this paper, we study growth and properties of *quaternary* alloy magnetic semiconductor $[(In_yGa_{1-y})_{1-x}Mn_x]As$ which is based on $In_yGa_{1-y}As$. This quaternary alloy system has many potential advantages. First, the bandgap energy can be changed over a wide range, roughly from 0.4 eV to 1.4 eV, by changing the In content $y$. Especially, the bandgap of (InGaMn)As can be tuned at the wavelength of 1.3 or 1.55 μm for optical communications, which cannot be realized by ternary alloy magnetic semiconductors. Another possibility is controllability of band structures and of easy magnetization axis (which is sensitive to the lattice strain) by changing the In content. Therefore, introduction of quaternary alloy magnetic semiconductors will broaden the possibility of bandgap engineering using semiconductor alloys and heterostructures. However, no paper about growth and properties of (InGaMn)As has been published so far.[8] In this paper, we present successful growth of ferromagnetic (InGaMn)As thin films by low-temperature MBE (LT-MBE).

We have grown 200 - 250 nm thick $[(In_{1-y}Ga_y)_{1-x}Mn_x]As$ films both on semi-insulating GaAs(001) substrates and on semi-insulating InP(001) substrates by LT-MBE. Typical growth procedure is as follows. When a GaAs substrate was used, after growing a 100 nm-thick GaAs buffer layer and a 0.6 μm thick $In_{1-y}Ga_yAs$ buffer layer successively at 500 °C, a 250 nm-thick $[(In_yGa_{1-y})_{1-x}Mn_x]As$ film was grown at 250 °C. When an InP substrate was used, after growing a 10 nm-thick $In_yGa_{1-y}As$ buffer layer at 500 °C, a 200 nm-thick $[(In_yGa_{1-y})_{1-x}Mn_x]As$ film was grown at 250 °C. The Mn content $x$ was varied from 0 to 0.08, and the In content $y$ was varied from 0 to 0.53. Reflection high energy electron diffraction (RHEED) patterns during the growth of (InGaMn)As were streaky with (1×$n$) reconstruction, as shown in Fig. 1.

When the electron beam azimuth was $[\bar{1}10]$, peculiar fractional order streaks were seen between integral order $(\bar{1}0)$, $(00)$, and $(10)$ streaks. The fractional order streaks are not exactly half-order. This peculiar reconstruction was observed for $[(In_yGa_{1-y})_{1-x}Mn_x]As$ with $0.2 = y = 0.53$.

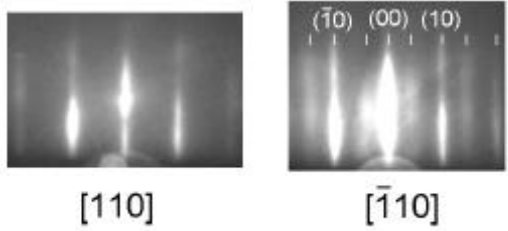

Fig. 1 RHEED patterns during MBE growth of $[(In_{0.26}Ga_{0.74})_{0.96}Mn_{0.04}]As$ at 250 °C with electron beam azimuths along [110] and [$\bar{1}$10].

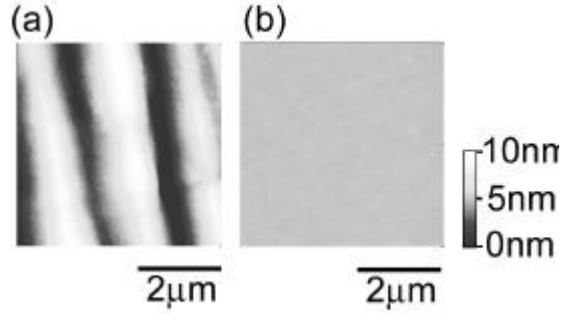

Fig. 2 Surface morphology taken by AFM of (a) a lattice mismatched $[(In_{0.26}Ga_{0.74})_{0.96}Mn_{0.04}]As$ thin film grown on a GaAs substrate and of (b) a nearly lattice matched $[(In_{0.53}Ga_{0.47})_{0.96}Mn_{0.04}]As$ thin film grown on an InP substrate. In (a) a very rough surface with many cross hatches was observed, while in (b) a very smooth surface with the average roughness of 0.37 nm (rms), corresponding to less than 2 monolayers, was observed.

For thin-film type devices to be integrated with III-V optoelectronics, the surface morphology of epitaxial films is very important. Figure 2 shows surface images of (a) a lattice *mismatched* $[(In_yGa_{1-y})_{1-x}Mn_x]As$ film ($x = 0.04$, $y = 0.26$) grown on a GaAs substrate and of (b) a *nearly lattice matched* $[(In_yGa_{1-y})_{1-x}Mn_x]As$ film ($x = 0.04$, $y = 0.53$) grown on an InP substrate, taken by atomic force microscopy (AFM). For the lattice mismatched (InGaMn)As film grown on GaAs, a very rough surface with many cross hatches with the roughness of 10 nm or more was obtained, as shown in Fig. 2(a). This means that this (InGaMn)As film grown on GaAs has many dislocations generated from the interface between the InGaAs buffer layer and the GaAs substrate. By contrast, for the nearly lattice matched (InGaMn)As film grown on InP, no crosshatches were observed, as shown in Fig. 2(b). The average surface roughness was only 0.37 nm (rms), corresponding to less than 2 monolayers. This morphology is almost the same as that of (GaMn)As films grown on GaAs. This result suggests that nearly lattice matched (InGaMn)As films grown on InP substrates, which have no dislocations and have very smooth surface morphology, are immediately applicable for thin-film type devices, while lattice mismatched (InGaMn)As films grown on GaAs substrates should be prepared more carefully to improve the surface morphology (For example, using pseudo-morphic ultra thin films or growing on a lattice-relaxed thick InGaAs buffer layer will be effective to suppress the surface roughness.)

In order to estimate the lattice constant *a* of (InGaMn)As, we have done X-ray diffraction measurements on 200 nm-thick $[(In_yGa_{1-y})_{1-x}Mn_x]As$ ($y = 0.53$) thin film samples grown on InP substrates whose Mn concentration $x$ was 0, 0.02, 0.04, 0.06 and 0.08. Here the In content $y$ was fixed at 0.53, thus the films are nearly lattice matched to InP. The intrinsic (unstrained) lattice constant *a* of (InGaMn)As was calculated from the diffraction peak in X-ray spectra, assuming that (InGaMn)As was fully strained and that (InGaMn)As has the same elastic constant as InGaAs. Figure 3 shows (a) X-ray diffraction spectra of $[(In_yGa_{1-y})_{1-x}Mn_x]As$ in the case of $x = 0.04$, $y = 0.53$, and (b) $x$ dependence of the lattice constant *a* of (InGaMn)As. The solid and broken lines in Fig. 3(a) are measured and simulated X-ray spectra, respectively. The broken line in Fig. 3(b) indicates the lattice constant derived by Vegard's law, where we used the value of 0.598 nm as the lattice constant of hypothetical zinc-blende MnAs.[3] From Fig. 3, we can see that the experimental lattice constant slightly decreases with increasing $x$ when $x = 0.04$. When $x$ reached 0.06, the lattice constant became close to the value predicted by Vegard's law. The reason why the lattice constant does not follow Vegard's law is not clear at present. There is no indication of MnAs clustering in the RHEED and X-ray results of all the samples examined here. One practically important point is that the lattice mismatch of $[(In_{0.53}Ga_{0.47})_{1-x}Mn_x]As$ films to InP is very small, only ± 0.1 %.

In order to study the magnetic and magneto-optical properties, we have measured magnetic circular dicroism (MCD) of $[(In_yGa_{1-y})_{1-x}Mn_x]As$ films with the Mn concentration $x$ fixed at 0.04 and the In content $y$ varied from 0 to 0.53. All the samples were found to be ferromagnetic when prepared under appropriate conditions. In some (not all) as-grown samples, we have seen no ferromagnetic order. However, after annealing these samples at 280 - 300 °C for 10 minutes, ferromagnetic ordering was obtained. This improvement of magnetic property by annealing is

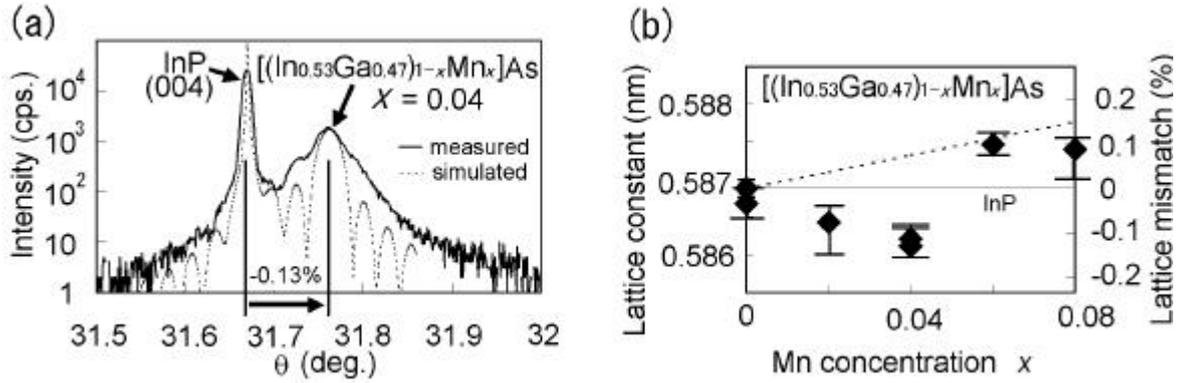

Fig. 3 (a) X-ray diffraction spectra of a 200 nm-thick [(In$_{0.53}$Ga$_{0.47}$)$_{0.96}$Mn$_{0.04}$]As film grown on an InP substrate. Solid and broken lines denote measured and simulated spectra, respectively. (b) Mn concentration $x$ dependence of the intrinsic lattice constant $a$ of [(In$_{0.53}$Ga$_{0.47}$)$_{1-x}$Mn$_x$]As calculated from the X-ray diffraction peak. The broken line indicates the value derived by Vegard's law, assuming the hypothetical zinc-blende MnAs lattice constant (0.598 nm, see the text). The horizontal solid line shows the lattice constant (0.58688 nm) of InP.

due to the increase of hole concentration caused by the reduction of excess As incorporated during the non-equilibrium low-temperature MBE growth.[9] Figure 4 shows MCD spectra of (a) [(In$_y$Ga$_{1-y}$)$_{1-x}$Mn$_x$]As ($x = 0.04$, $y = 0.26$) measured at 8 K and of (b) (GaMn)As as a reference measured at 5 K, in a reflection set up with a magnetic field of 1 T applied perpendicular to the film. Both samples were grown on GaAs substrates. The result of [(In$_y$Ga$_{1-y}$)$_{1-x}$Mn$_x$]As ($x = 0.04$, $y = 0.53$) grown on InP was similar (not shown here). For (InGaMn)As, strong negative peaks were observed at 1.46 eV (near $E_0$) and at 2.7 - 2.8 eV (near $E_1$). Here, $E_0$ and $E_1$ correspond to the optical transitions at G and at ? critical points, respectively. The MCD intensity of (InGaMn)As was very strong, and comparable to or even stronger than that of (GaMn)As.[10] The spectral features of (InGaMn)As were similar to those of (GaMn)As, indicating that the band structure of (InGaMn)As is of zinc-blende semiconductor type. It was seen from the spectra that $E_0$ and $E_1$ critical point energy values were shifted to lower energy due to the reduction of bandgap. For example, in Fig. 4 (a) and (b) the largest negative peak energy coming from $E_0$ shifted from 1.62 eV to 1.46 eV. The amount of this peak energy shift in MCD was 0.16 eV, which is smaller than the $E_0$ shift ~0.36 eV expected from the bandgap reduction by adding the same content of In ($y = 0.26$) to GaAs. In MCD spectra a peak does not directly indicate a critical point energy but a value of energy *near* the critical point. In this experiment, we measured reflection MCD whose peaks do not exactly indicate critical points, and also the peak energy depends on the carrier concentration especially in the heavily doped semiconductors such as (GaMn)As and (InGaMn)As. This red shift of the MCD peak, however, was observed in all the (InGaMn)As samples and tend to increase with increasing the In content. This result indicates that the bandgap energy can be changed in principle by changing the In content of (InGaMn)As films.

Furthermore, we measured magnetic field dependence of the MCD intensity of this (InGaMn)As film at 1.46 eV at various temperatures, as shown in Fig. 4 (c). The magnetic field was applied perpendicular to the film plane, thus the MCD signal is proportional to the perpendicular component of magnetization. We can see a clear hysteresis loop, indicating a ferromagnetic order up to 30 K. The reason why the Curie temperature of this (InGaMn)As sample is lower than those of (InMn)As and (GaMn)As is probably that the growth condition of (InGaMn)As is not fully optimized and there are many excess As in the (InGaMn)As lattice. We expect that higher Curie temperature will be obtained by increasing the growth temperature or by reducing the As pressure during the growth. It was also found that the easy magnetization axis of the (InGaMn)As film is perpendicular to the film plane. This is consistent with the fact that this [(In$_y$Ga$_{1-y}$)$_{1-x}$Mn$_x$]As ($y = 0.26$, $x = 0.04$) film receives tensile strain and its lattice constant is smaller than that of In$_y$Ga$_{1-y}$As ($y = 0.26$), as in the case of $y = 0.53$ described earlier. Also this result indicates that the easy magnetization axis is controllable by changing the strain generated in the (InGaMn)As layer by changing the Mn concentration $x$ or In content $y$.

In summary, ferromagnetic [(In$_y$Ga$_{1-y}$)$_{1-x}$Mn$_x$]As quaternary alloy thin films were successfully grown by LT-MBE both on GaAs (001) substrates and on InP(001) substrates. The surface morphology of (InGaMn)As can be very smooth when nearly lattice matched films were grown on InP substrates. The lattice constants of [(In$_y$Ga$_{1-y}$)$_{1-x}$Mn$_x$]As thin films were slightly smaller (less than 0.2 %) than that of In$_y$Ga$_{1-y}$As with the same In content $y$ when the Mn concentration $x$ is less than 0.04.  MCD studies show that (InGaMn)As has zinc-blende type band structure, strong magneto-optical effect at $E_0$ and $E_1$ critical point energies, and perpendicular magnetization due to the tensile strain.

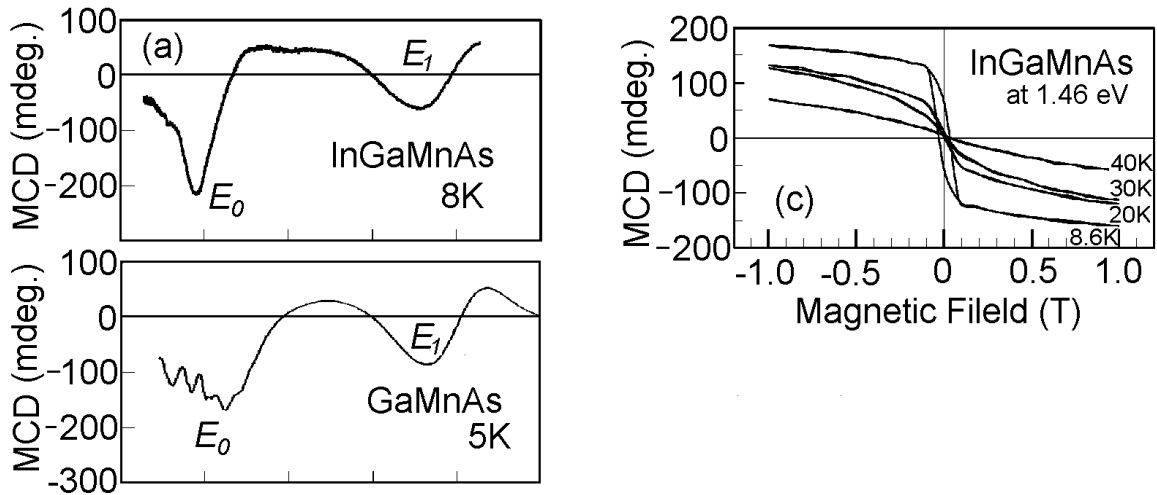

Fig. 4  MCD spectra of (a) [(In$_{0.26}$Ga$_{0.74}$)$_{0.96}$Mn$_{0.04}$]As grown on a GaAs substrate at 8 K, and of (b) (GaMn)As at 5 K (for reference).  (c) Magnetic field dependence of MCD measured on (InGaMn)As (same sample as (a)) at 8.6 K, 20 K, 30 K and 40 K.  All the MCD measurements were done in a reflection set up, under a magnetic field of 1 T applied perpendicular to the substrate.

Acknowledgments
 The authors thank Prof. T. Nishinaga and Prof. S. Naritsuka of Meijo University for their constant support.  Thanks are also due to Dr. S. Sugahara for his collaboration and help.  This work was partially supported by the JSPS Research for the Future Program (JSPS-RFTF97P00202), by the CREST project of JST, and by Grant-in-Aid for Scientific Research from the Ministry of Education, Science, Sports and Culture.